\documentstyle[12pt,aps,epsfig]{revtex}

\begin{document}
\begin{flushright}
\baselineskip=24pt

{\bf 
AS-ITP-98-13\\
MADPH-98-1089\\
hep-ph/9810412}\\

\end{flushright}

\begin{center}
\vglue 0.5cm
{\Large\bf  Scales, Couplings Revisited and Low Energy Phenomenology
in M-theory on $S^1/Z_2$\\}
\vglue 1.0cm
{\Large Chao-Shang Huang$^1$, Tianjun Li$^2$, 
Wei Liao$^1$, Qi-Shu Yan$^1$ }
\vglue 0.5cm
\begin{flushleft}
$^1$Institute of Theoretical Physics, Academia Sinica, P. O. Box
2735, Beijing 100080, P. R. China \\
$^2$Department of Physics, University of Wisconsin, Madison,
WI 53706,
U. S. A. \\
\end{flushleft}
\vglue 0.2cm
{\Large 
 Shou Hua Zhu}
\vglue 0.7cm
\begin{flushleft}
CCAST (World Lab), P.O. Box 8730, Beijing 100080, P.R. China \\
and Institute of Theoretical Physics, Academia Sinica, P. O. Box
2735,\\ Beijing 100080, P. R. China \\
\end{flushleft}
\end{center}

\vglue 0.3cm
\begin{center}
{\bf ABSTRACT}
\end{center}
\begin{abstract}
We revisit the eleven dimension Planck scale, the
physical scale of the eleventh dimension, the 
physical scale of  
Calabi-Yau manifold 
 and coupling in hidden sector in M-theory on
$S^1/Z_2$. And we discuss the reasonable bound on them.
Considering F-term of dilaton and moduli SUSY breaking
and choosing two repersentative points which 
correspond to scalar quasi-massless scenario and
dilaton dominant SUSY breaking scenario
respectively,  we analyze experimental constraints to the parameter space.
 The sparticle spectrum and some phenomenological 
predictions are also given. 
\end{abstract}

PACS:supersymmetric models, 12.60.Jv

\newpage
\setcounter{page}{1}
\pagestyle{plain}
\baselineskip=24pt

\section{Introduction}
In recent years revolutional progress in our understanding of string theories
has been made. The key discoveries were dualities, which show that the five
distinct superstring theories are in fact five different perturbative
expansions of a single underlying theory (11-dimensional M theory or 
12-dimensional F theory) about five different points in the moduli space of
consistent vacua. The dualities, furthermore, show that in addition to the 
five points of the moduli space, there is a sixth special point in the 
moduli space which involves an 11-dimensional Minkowski space-time and is
related to the strongly coupled heterotic (HE) and IIA superstring theories
by compactifications on $S^1$/$Z_2$ and $S^1$ respectively~\cite{RV}. Nowadays 
we do 
not have the complete picture of M theory so that one might argue that it
is premature to make any attempt at phenomenology. However, experiences of
investigating weak coupled $E_8 \times E_8'$ HE superstring phenomenology
tell us that it may be that the corners of the moduli space capture most of 
the features of the theory relevant for low energy phenomenology.

Since Horava and Witten~\cite{HW} described the strongly coupled $E_8\times E'_8$
HE string theory by M-theory compactified on $S^1/Z_2$ whose
low energy limit is the eleven-dimensional supergravity, many interesting
phenomenology implications have been studied: Newton's constant 
and compactification,
gluino condensation and supersymmetry breaking, Axions and Strong CP
problem, threshold scale and strong coupling effects, 
proton decay, and phenomenological
consequences~\cite{Witten,Horava,BD,KC,AQ,AQR,EDCJ,VK,LLN,TIAN,BL,JUN,NOY,LT,LOD,LOW,CKM,BKL,EFN,DM,KARIM,SST,NSLPT,NSLOW}
 ( for a review, see
Ref.~\cite{DVN}). In short, all of above result seems to show that M-theory
is a better candidate than previous weak-coupled heterotic string theory.

The most important one of discoveries of M theory phenomenology is that 
the discrepancy between the Grand Unification scale of around $2\times 10^
{16}$ GeV estimated by extrapolating from the LEP measurements and the
estimate of around $4\times 10^{17}$ GeV calculated in the weak coupled
$E_8\times E'_8$ HE string theory may be removed in the strongly coupled
$E_8\times E'_8$ HE string theory. In Horava and Witten's picture, at one
end of the 11th dimensional line segment of length $\pi\rho$ live the 
observable fields contained in $E_8$, at the other end live the hidden
sector fields contained in $E'_8$, and in the middle ('bulk") propagate
the gravitational fields. One needs to consider further $R_4\otimes X_{CY}$
($X_{CY}$ denotes a 6-dimensional Calabi-Yau manifold) compactification
of 10-dimensional $E_8\times E'_8$ HE string in order to get a realistic
effective theory. Therefore,  there are several scales and couplings such
as the eleven dimension Planck scale, the physical 
scale of the eleventh dimension, the physical scale of  
Calabi-Yau manifold and the couplings in the observable and hidden sectors.
The values of these scales and couplings and the relations between them
have been estimated~\cite{Witten,TIAN,NOY}.

Because the values of the scales and couplings are important for 
phenomenology and there are some issues which need to be discussed,
In this paper we shall first revisit the eleven dimension Planck scale, 
the physical scale of the eleventh dimension, the physical scale of  
Calabi-Yau manifold and the coupling ( in terms of the 
function x which is defined by 
 $x = {{\alpha_H \alpha_{GUT}^{-1} - 1} \over\displaystyle {\alpha_H \alpha_{GUT}^{-1} + 1}} $)
  in hidden sector in M-theory on $S^1/Z_2$
in standard embedding and non-standard embedding~\cite{KARIM,SST,NSLPT,NSLOW}, 
and then discuss the possible bounds on them from
the ansatz that the eleven-dimension Planck scale is
larger than the $M_{GUT}$, $M_H$ which is the scale in the hidden sector
just after the Calabi-Yau manifold is compactified,
the eleventh dimension scale $ \left[\pi \rho_p\right]^{-1} $. For the 
standard embedding, we obtain the upper bound on
x is 0.97 ( $x < 0.97 $ ),
for $\alpha_{GUT} = {1\over 25}$. 

An important scale which is directly relevant to phenomenology is the scale
$\Lambda_{SUSY}$ from which the soft terms start running. There is a significant
difference for $\Lambda_{SUSY}$ between the weakly coupled and strongly
coupled limits. In the weakly coupled limit $\Lambda_{SUSY}$ is close to
$M_{Pl}$ since observable and hidden sector fields as well as gravitational
fields all live in a same 10-dimensional space-time. In the strongly coupled
limit, as Horava~\cite{Horava} has argued that SUSY breaking is not felt immediately
in the observable sector because of a topological obstruction (the 11th
dimension separates the two sectors). SUSY breaking in the hidden sector
communicates to the observable sector by gravitational interactions. Therefore,
SUSY breaking in the observable sector becomes apparent only after the
renormalization scale Q is low enough to not reveal the presence of the 11th
dimension anymore. Therefore, a natural and reasonable choice is
 $\Lambda_{SUSY}=\left[\pi \rho_p\right]^{-1}$. We estimate the value of
 $ \left[\pi \rho_p\right]^{-1} $ and get its low bound of 9.5$\times 10^{13}
$ GeV.

In order to discuss the M-theory low energy phenomenology, we may need to
pay attention to the supersymmetry
breaking in M-theory and think about M-theory model building (essentially
 compactifications of 6-dimensional space-time),
which is similar to what happened 10 years ago. As we know,
we can discuss the supersymmetry breaking in the following
ways: non-zero F-terms of the dilaton or moduli fields SUSY breaking
in which we do not specify the trigger of the SUSY breaking
~\cite{NOY,LOW,CKM},  and 
the Scherk-Schwarz mechanism on the 
eleventh ( or fifth ) dimension ( or we might call it 
coordinate-dependent compactification)~\cite{AQR,EDCJ}.
In this paper we consider the phenomenology in
non-zero F-terms of the dilaton and/or moduli SUSY breaking. 
>From the phenomenological view, the
important features of M theory phenomenology which are different from the
weakly coupled limit and independent of the details of M theory model
building are of unification of couplings and the magnitude of $\Lambda_{SUSY}
$ and the emphases of this paper are investigating characteristic features
of low energy phenomenology of M theory. In this paper, we take the simplest
compactification as an example (like most of people did)
and choose two representative points which 
correspond to scalar quasi-massless scenario and
dilaton dominant SUSY breaking scenario
respectively. Then we calculate the low-energy sparticle spectrum under the LEP
experiment constraints and discuss its dependence  on $\Lambda_{SUSY}$.
It is found that $M_{1/2}$ can not be larger than 400 GeV if one demands
that masses of sparticles are not beyond 1 TeV. We analyze the constraints
to the parameter space from $b\rightarrow s\gamma$. It is found that in the
dilaton dominant SUSY breaking scenario although $b\rightarrow s\gamma$
imposes stringent constraints to the parameter space there still is a region
of the parameter space where tan$\beta$ is large and $M_{1/2}$ is small,
which will lead to significant SUSY effects in some processes.

In this paper, we discuss the scales and couplings  in the
section 2. In section 3, we discuss soft terms. In section 4 we calculate
sparticle spectrum using revised ISAJET. Section 5 is devoted to analyze
the constraints from $b\rightarrow s\gamma$. We discuss the rare decay
$B\rightarrow X_s \tau^+\tau^-$ and search of Higgs bosons in section 6.
Finally, section 7 contains our conclusion.

\section{ Eleventh Dimension Scale and  Gauge Coupling
in Hidden Sector.}
First, let's consider the gauge couplings, gravitational
coupling and the physical eleventh dimension
 radius in the M-theory. The relative
11-dimensional Lagrangian is given by~\cite{HW}
\begin{eqnarray}
L_B&=&-{1\over {2 \kappa^2}}\int_{M^{11}}d^{11}x\sqrt g
R - \sum_{i=1,2}
{1\over\displaystyle 2\pi (4\pi \kappa^2)^{2\over 3}}
\int_{M^{10}_i}d^{10}x\sqrt g {1\over 4}F_{AB}^aF^{aAB} ~.~\,
\end{eqnarray}
In the 11-dimensional 
metric~\footnote{Because we think 11-dimension
metric is more fundamental than string metric
and Einstein frame, 
 we discuss the scales and couplings in 11-dimension metric.}, 
the gauge coupling and gravitational
coupling in 4-dimension are~\cite{Witten,TIAN}:
\begin{eqnarray}
8\pi\,G_{N}^{(4)} &=& {\kappa^2 \over 
{2\pi \rho_p V_p}} ~,~ \, 
\end{eqnarray}
\begin{eqnarray}
\alpha_{\rm GUT} &=&{1\over {2 V_p (1+x)}}\,(4\pi\kappa^2 
)^{2/3} ~,~ \,
\end{eqnarray}
\begin{eqnarray}
\left[\alpha_H \right]_W &=&{1\over {2 V_p (1-x)}}\,(4\pi\kappa^2 
)^{2/3} ~,~\,
\end{eqnarray}
where $x$ is defined by:
\begin{eqnarray}
x &=& \pi^2 {\rho_p \over V_p^{2/3}} 
({\kappa \over 4 \pi })^{2/3} \int_X \omega \wedge 
{{trF \wedge F - {1\over 2} tr R \wedge R}
\over\displaystyle { 8 \pi^2}} ~,~\,
\end{eqnarray}
where $\rho_p$, $V_p$ are the physical eleventh dimension radius
and Calabi-Yau manifold volume ( which is defined by the middle 
point Calabi-Yau manifold 
volume between the observable sector and the hidden sector )
respectively, and $V_p = V e^{3\sigma}$
where $V$ is the internal Calabi-Yau volume (For detail, 
see Ref.~\cite{TIAN}).
>From above formula, we can obtain that:
\begin{eqnarray}
x &=& {{\alpha_H \alpha_{GUT}^{-1} - 1} \over\displaystyle
{\alpha_H \alpha_{GUT}^{-1} + 1}} ~.~\,
\end{eqnarray}
The GUT scale $M_{GUT}$ and  the hidden sector scale
$M_H$ when 
the Calabi-Yau manifold is compactified are:
\begin{eqnarray}
M_{\rm GUT}^{-6} &=& V_p ( 1+x) ~,~\,
\end{eqnarray}
\begin{eqnarray}
M_H^{-6} &=& V_p ( 1-x) ~,~\,
\end{eqnarray}
or we can express the $M_H$ as:
\begin{eqnarray}
M_H &=& ({\alpha_H \over\displaystyle 
\alpha_{GUT}})^{1/6} M_{GUT} = ({{1+x} 
\over\displaystyle {1-x}})^{1/6} M_{GUT} ~.~\,
\end{eqnarray}
Noticing that $M_{11} = \kappa^{-2/9}$, we have
\begin{eqnarray}
M_{11} &=& \left[2 (4\pi )^{-2/3}\,  
\, \alpha_{\rm GUT}\right]^{-1/6} M_{GUT}  ~.~\,
\end{eqnarray}
And we can also obtain the physical 
 scale of the eleventh dimension in
the eleven-dimensional metric:
\begin{eqnarray}
\left[\pi \rho_p\right]^{-1} &=& 
{{8 \pi}\over\displaystyle {1+x}} \left(2 
\alpha_{\rm GUT}\right)^{-3/2}
{{M_{GUT}^3}\over\displaystyle
  {M_{Pl}^2}}~.~\, 
\end{eqnarray}
Now, we consider constraints. Our ansatz is that
the scale of $M_{GUT}$, $M_H$ and 
$\left[\pi \rho_p\right]^{-1}$  should be lower than the
eleven dimension Planck scale. From the
constraints $M_{GUT}$ and $M_H$ is smaller than
the scale of $M_{11}$, we obtain that:
\begin{eqnarray}    
\alpha_{GUT} \leq {{(4 \pi)^{2/3}}\over\displaystyle
2} ~;~
\alpha_H \leq {{(4 \pi)^{2/3}}\over\displaystyle
2} ~,~ \,
\end{eqnarray}
or 
\begin{eqnarray}    
\alpha_{GUT} \leq 2.7 ~;~
\alpha_H \leq 2.7 ~,~ \,
\end{eqnarray}
they are independent number and
large enough for our discussion. For the 
standard embedding, we obtain the upper bound on
x is 0.97 ( $x < 0.97 $ ),
for $\alpha_{GUT} = {1\over 25}$. 
>From the constraints 
that $ \left[\pi \rho_p\right]^{-1} $ is smaller
than the scale of $M_{11}$, we obtain that:
\begin{eqnarray} 
M_{GUT} \alpha_{GUT}^{-2/3} \leq 
\sqrt {1+x} 2^{1/6} (4 \pi )^{-4/9}
M_{Pl} ~,~ \,
\end{eqnarray}
which is obviously satisfied for standard embedding.
However, if we
 can consider non-standard embedding $ x < 0 $
 ~\cite{KARIM,SST,NSLPT,NSLOW}, i. e., 
the gauge coupling in the observable sector is larger
than the coupling in the hidden sector, we will have 
the following low bound on x:
\begin{eqnarray} 
x_{lb} \geq 2^{-1/3} (4 \pi )^{8/9} 
(\alpha_{GUT})^{-4/3} 
{{M_{GUT}^2} \over\displaystyle {M_{Pl}^2}} -1 ~.~ \,
\end{eqnarray}
Therefore, there exist three possibilities between the
physical scale of the eleventh dimension and the
physical scale of the Calabi-Yau manifold: 
$ \left[\pi \rho_p\right]^{-1} $ is smaller than
$M_{GUT}$ and $M_H$ which, from low energy
to high energy,   corresponds to from 4-dimension to
5 dimension and then, to 11-dimension;  
$ \left[\pi \rho_p\right]^{-1} $ is smaller than
$M_{GUT}$ but larger than $M_H$, which,
 assuming $x^{11}$ is the
coordinate of the eleventh dimension, and the observable
sector is at $x^{11}=0$ plane and the hidden sector at
$x^{11} = \int dx^{11} \sqrt {g_{11, 11}}$ or the opposite plane, 
from low energy to high energy,
corresponds to at one 
particular point $x^{11}_c$, from 4 dimension to 11 dimension
directly, for $x^{11} < x^{11}_c$,  from 4-dimension to
5 dimension and then, to 11-dimension, and for 
$x^{11} > x^{11}_c$,  from 4-dimension to
10 dimension and then, to 11-dimension;
$ \left[\pi \rho_p\right]^{-1} $ is larger than
$M_{GUT}$ and $M_H$ which, from low energy
to high energy,   corresponds to from 4-dimension to
10 dimension and then, to 11-dimension.
Let us define the $x_H$ and $x_O$ which correspond
to $ \left[\pi \rho_p\right]^{-1} = M_H $ and 
$  \left[\pi \rho_p\right]^{-1} = M_{GUT} $
respectively. 
\begin{eqnarray} 
\left[{{(1+x_H)^7}\over\displaystyle 1-x_H}\right]^{1/6}  
&=& 8 \pi  
(2 \alpha_{GUT})^{-3/2} 
{{M_{GUT}^2} \over\displaystyle {M_{Pl}^2}} ~,~ \,
\end{eqnarray} 
\begin{eqnarray} 
x_O &=& 8 \pi  
(2 \alpha_{GUT})^{-3/2} 
{{M_{GUT}^2} \over\displaystyle {M_{Pl}^2}} -1 ~.~ \,
\end{eqnarray} 
It is
obvious that from eq. (11) when $x$ decreases,  
$ \left[\pi \rho_p\right]^{-1}$ increases if we consider specific
$\alpha_{GUT}$ and $M_{GUT}$, so, we have 
$x_H \geq x_O \geq x_{lb}$.

Now we can discuss the numerical result. We take 
$M_{GUT} =2.0 \times 10^{16} $ GeV, 
$ \alpha_{GUT} ={ 1 \over{25}} $, $M_{Pl}
=2.4 \times 10^{18} $ GeV, then, we obtain the
$M_{11} =4.04 \times 10^{16} $ GeV, 
$x_{lb} = -0.96 $, $x_O = -0.92$,
$x_H=-0.878$,
$ \left[\pi \rho_p\right]^{-1} $ is from 
$7.8 \times 10^{14}$ GeV to 
$1.5 \times 10^{15}$ GeV when we vary x
from 0.97 to 0 in the mean time
for the standard embedding.
If we choose the $M_{GUT}$ is $3 \times 10^{16}$ GeV,
we obtain
$M_{11} =6.05 \times 10^{16} $ GeV, 
$x_{lb} = -0.91 $, $x_O=-0.826$, $x_H = -0.758$,
$ \left[\pi \rho_p\right]^{-1} $ is from 
$2.64 \times 10^{15}$ GeV to 
$5.2 \times 10^{15}$ GeV when we vary x 
from 0.97 to 0 in the mean time.
And we notice that $x_{lb}$, 
$x_O$, $x_H$ increase if we increase the $M_{GUT}$. 
Therefore, if we had large GUT scale 
because of additional matter fields in
the future M-theory model building,
we might need to pay attention to $x_{lb}$,  
$x_O$, $x_H$ in order to get clear picture of the universe.

Furthermore, we can discuss the possible low energy scale
of $ \left[\pi \rho_p\right]^{-1} $ which is interesting
for the low energy phenomenology when $x > 0$
for standard embedding. Let us 
define the
relation between the physical Calabi-Yau manifold
volume and the unification scale $M_{GUT}$ as in 
~\cite{NOY}:
\begin{eqnarray}
 a M_{\rm GUT}^{-1} &=& (V_p ( 1+x))^{1/6} ~.~\,
\end{eqnarray} 
where $a > 1$. And a is smaller than 
2.02 in order to keep the $M_{GUT} < M_{11}$   if
we take $ \alpha_{GUT} ={ 1 \over{25}} $. The formula
is similar to above except the transformation:
$M_{GUT} \rightarrow {M_{GUT} \over a} $.
Taking  $M_{GUT} =2.0 \times 10^{16} $ GeV, 
$ \alpha_{GUT} ={ 1 \over{25}} $, $M_{Pl}
=2.4 \times 10^{18} $ GeV, we get the low bound on 
$ \left[\pi \rho_p\right]^{-1} $ is 
$9.5 \times 10^{13} GeV$. Using 
 $M_{GUT} =3.0 \times 10^{16} $ GeV, the low bound
 is $3.2 \times 10^{14}$ GeV. It follows that $\Lambda_{SUSY}\geq 10^{14}$ GeV, which
is consistent with the estimate given in the Ref.\cite{NOY}. 
 
\section{Soft terms} 

The k\"ahler potential, gauge kinetic function and
the superpotential in the simplest compactification
of M-theory on $S^1/Z_2$ are ~\cite{NOY,LOD}
~\footnote{We choose this simplest case as an example. In fact,
 if we consider three families and three moduli,
 in order to avoid FCNC problems that might arise from the violation
of the universal scalar masses in three families ( although this
kind of the violation might be very small),  we might need to assume
 that:
$\alpha_1 (T_1 + \bar T_1 ) = \alpha_2 (T_2 + \bar T_2 )
=  \alpha_3 (T_3 + \bar T_3 )$, 
and
$F^{T_1}=F^{T_2}=F^{T_3}$
where $\alpha_i$ i=1, 2, 3 are the next order correction constants.
Then, the final soft terms will be the same as
the simplest case. 
so, it is reasonable to choose the simplest
case as an example to analyze the phenomenology.}:
\begin{eqnarray}
K &=& \hat K + \tilde K |C|^2 ~,~ \,
\end{eqnarray}
\begin{eqnarray}
\hat K &=&  -\ln\,[S+\bar S]-3\ln\,[T+\bar T] ~,~\,
\end{eqnarray}
\begin{eqnarray}
\tilde K &=& ({3\over\displaystyle {T+\bar T}} +
{\alpha\over\displaystyle {S+\bar S}}) |C|^2  ~,~ \,
\end{eqnarray}
\begin{eqnarray}
Ref^O_{\alpha \beta} &=& Re(S + \alpha T)\, \delta_{\alpha \beta} ~,~\,
\end{eqnarray}
\begin{eqnarray}
Ref^H_{\alpha \beta} &=& Re(S - \alpha T)\, \delta_{\alpha \beta} ~,~\,
\end{eqnarray}
\begin{eqnarray}
W= d_{x y z} C^x C^y C^z ~,~\,
\end{eqnarray}
where $S$, $T$ and $C$ are dilaton, moduli and matter fields 
respectively. $\alpha$ is a next order correction constant 
which is related to the  Calabi-Yau manifold.

With those information, we have following 
soft terms\cite{JUN,CKM}:
\begin{eqnarray}
M_{1/2}&=&{{\sqrt 3  M_{3/2}} \over\displaystyle {1+x}}
(sin\theta +{x\over \sqrt 3} cos\theta ) ~,~\,
\end{eqnarray}
\begin{eqnarray}
M_0^2&=& M_{3/2}^2 - 
{{3  M_{3/2}^2} \over\displaystyle 
{(3+x)^2}}(x(6+x) sin^2\theta +
\nonumber\\&&
 ( 3 + 2x ) cos^2\theta
- 2 \sqrt 3 x ~sin\theta ~cos\theta) ~,~ \,
\end{eqnarray}
\begin{eqnarray}
A&=&- {{\sqrt 3  M_{3/2}} \over\displaystyle 
{(3+x)}} ((3-2x) sin\theta + \sqrt 3~ x ~cos\theta) ~,~ \,
\end{eqnarray}
where $M_{3/2}$ is the gravitino
mass, the quantity x defined above can be also expressed as
\begin{eqnarray}
x={{\alpha ( T + \bar T )} \over\displaystyle { S + \bar S }} 
 ~.~\,
\end{eqnarray}

We pick the following two points as representatives
which correspond to scalar quasi-massless and the dilaton dominant
scenario.
The soft terms and parameters for the first point are:
\begin{eqnarray}
M_{1/2}=0.989 M_{3/2} ~,~ M_0 = 0.008 M_{3/2}~,~  \,
\end{eqnarray}
\begin{eqnarray}
A=-0.761 M_{3/2}~,~ x=0.5838, ~,~ \tan\theta=-4.566 ~.~ \,
\end{eqnarray}
and the soft terms and parameters for the second are: 
\begin{eqnarray}
M_{1/2}= 1.534 M_{3/2}~,~ M_0 = 0.870 M_{3/2} ~,~  \,
\end{eqnarray}
\begin{eqnarray}
A=-1.517 M_{3/2}~,~ x=0.13~,~ \theta={\pi \over 2}~.~ \,
\end{eqnarray}

\section{Mass spectra and the permitted parameter space}

We concentrate on the two typical supersymmetry breaking (SB) scenarios given in section 3
to calculate the low energy spectrum of superpartner and Higgs bosons masses: the scalar
quasi-massless scenario corresponding to $m_0=8.09 \times 10^{-4} M_{1/2}$ and 
$A = -0.769 M_{1/2}$ (see eqs.(32),(33)) the dilaton dominant scenario corresponding to $m_0=0.567 M_{1/2}$ 
and $A= -0.989 M_{1/2}$ (see eqs.(34),(35)). In order to find out the effects of the supersymmetry breaking scales to low
energy phenomenology, we take supersymmetry breaking scales $\Lambda_{SUSY}$ as $2.0 \times 10^{16}$ GeV
(the GUT scale), $1 \times 10^{15}$ GeV, and
$1 \times 10^{14}$ GeV. Those scales lower than $1\times10^{14}$ GeV are not chosen because of the analysis in Section 2. But we will discuss their possible effects also. 

We require that the lightest neutralino be the lightest supersymmetric particle (LSP) and 
use several experimental limits to constraint the parameter space, including 1)the 
width of the decay $Z \rightarrow \chi^0_1
\chi^0_1$ is less than 8.4 MeV, and branching ratios of $Z \rightarrow \chi^0_1
\chi^0_2$ and $Z \rightarrow \chi^0_2 \chi^0_2$ are less than 
$2 \times 10^{-5}$, where $\chi^0_1$ is the lightest neutralino and $\chi^0_2$ is
the other neutralino, 2) the mass of light
neutral even Higgs can not be lower than 77.7 GeV as the present experments
required, 3) the mass of lighter chargino must be larger than 65.7 GeV as given
by the Particle Data Group~\cite{PDG}, 4) sneutrinos are larger than 43.1 GeV,
5) seletrons are larger than 58.0 GeV, 6) smuons larger than 55.6 GeV, 7)
staus larger than 45.0 GeV.

We use ISAJET to do numerical calculations. In order to include all effects of bottom
and tau Yukawa couplings, we made some modifications to ISAJET which are the same as
those in Ref.\cite{TT}. We first examine the $M_{1/2}$ dependence of sparticle and Higgs
boson masses in the two SUSY breaking scenarios. It is found that the masses increase
when $M_{1/2}$ increase and $M_{1/2}$ should not
be larger than 400 GeV if one demand the masses of superpartner and Higgs bosons are
below 1 TeV. Then we scan boundaries of the parameter space in the two scenarios,
taking $M_{1/2}$ from zero to 400 GeV. For a certain scenario, there are only two
free parameters, $M_{1/2}$ and tan$\beta$, as well as sign($\mu$) under the radiative
electroweak symmetry breaking mechanism. The boundaries of the plane of the two 
parameters will be determined by the consistent conditions, such as the input should 
naturally trigger electric-weak symmetry breaking, the gauge unification,
the Yukawa couplings are in the perturbative range,
and there should be no tachyonic particles in mass spectrum, and by experimental
limits above listed. 
 
The results are shown in figure 1. The curves in figure 1 represent 
the upper bound and lower bound of tan$\beta$ for each $M_{1/2}$,
for two different SB scenarios, different $\Lambda_{SUSY}$. Figure 1a) draws the 
boundaries for $\mu < 0$~, figure 1b) for $\mu > 0$.
The dotted line represents $\Lambda_{SUSY}$ equals $10^{14}$ GeV, the dashed line 
$10^{15}$ GeV, and the solid line the GUT scale, $2.0 \times 10^{16}$ GeV. The curves
marked (1) (2) are the boundaries of the parameter spaces in the quasi-massless scenario
(the dilaton dominant scenario). For the 
scalar quasi-massless scenario, the permitted parameter spaces are the areas enclosed
with closed boundaries, while for the dilaton dominant scenario, the
permitted parameter spaces are not closed in right parts. The lower boundaries of 
tan$\beta$ is about 1.6 for the dilaton dominant scenario.
It is obvious from the figure 1 that for the scalar quasi-massless 
scenario, the parameter space is tightly constrained to the low mass spectrum and
no large tan$\beta$ region by consistent conditions,
which is similar to that found in Refs. \cite{LLN} and \cite{BKL}, while for the dilaton 
dominant scenario there is a much larger parameter space allowed.

The effect of sign of $\mu$ to the permitted parameter space
is significant, as can be seen by comparing figure 1a) and figure 1b). For example,
in the scalar quasi-massless scenario and  $\Lambda_{SUSY}$=$10^{14}$ GeV, if $\mu < 0$,
the parameter space is completely
excluded. While as $\mu > 0$, there does exist an allowed region.
The shape of boundaries of the parameter space for different sign of $\mu$ also
claims the effect. But the effect in the dilaton dominant scenario is not as sensitive
 as in the quasi-massless scenario.

It is interesting that for the case of dilaton dominant scenario and $\mu < 0$, the 
lower boundary of tan$\beta$ is singlely determined by the experimental limit of light 
Higgs mass. If the limit increases, the lower boundary will increase correspondingly.
It may be  understood from the tree level formula of the mass of light Higgs.
While the upper bound of tan$\beta$ is determined by both some experimental limits and LSP condition.
For example, when $\Lambda_{SUSY}$ is $10^{14}$ GeV, $M_{1/2}$ from 84.7 GeV to 88.2 GeV,
the upper bound is determined by the requirement that the mass of
$M_z$ should be less than $2m_{\tilde {u}_l}$, $2m_{\tilde {e}_l}$, $2m_{\tilde {e}_r}
$, $2m_{\tilde {\tau}_1}$, $2m_{\tilde {b}_1}$, and $2m_{\tilde {t}_1}$; from 88.2 GeV
to 102.0 GeV, it is determined by $m_{\chi_1^{\pm}} > 65.7$ GeV; from
102.0 GeV to 400 GeV, by LSP condition.

An interesting aspects of the allowed parameter space of the dilaton dominant scenario
is that there exists a region where the mass spectrum is low while tan$\beta$
is large. From figure 1, one obtains that the region increases when $\Lambda_{SUSY}$
decreases in the case of $\mu < 0$ and vice versa in the case of $\mu > 0$.
We find that if the limit of 
the lightest chargino mass increases, the region will be reduced. We know $b \rightarrow
s \gamma$ puts a very stringent constraint upon parameter space of MSSM.
In the region, the charged Higgs mass is about 150 GeV and consequently it will lead
to a significant contribution to $b\rightarrow s\gamma$. Therefore we would like to 
ask whether such a region can pass the constraint of $b \rightarrow s \gamma$. We
will answer this question in the next section.

We illustrate the tan$\beta$ dependence of mass spectra in the dilaton dominant 
scenario in figure 2, where we have chosen $M_{1/2}$=120 GeV. We have chosen this
value of $M_{1/2}$ because it is in the region of the parameter space pointed out above
and, as noticed in Ref.\cite{TT}, a study of this point serves to nicely illustrate the 
importance of large tan$\beta$ effects on Tevatron signals.
Spectra are drawn in the same graph for $\mu > 0$ and $\mu < 0$ denoted by solid and
dashed lines respectively. In figure 2a), the
supersymmetry breaking scale is $10^{14}$ GeV, while in figure 2b), the scale is
 $1\times10^{16}$ GeV. It is apparent that the mass spectrum will
drop with the decrease of $\Lambda_{SUSY}$, just as given in Ref. \cite{LLN}, because
the mass spectra depend on the length of the running scale of soft terms. The shorter 
the length, the lower the mass spectrum. This relation between the length of running
scale and mass spectra will keep till the $\Lambda_{SUSY}$ is lower than $10^9$ GeV. 
It is evident from figure 2 that the sign of $\mu$ can effect the spectrum, 
though not significantly.
It is also manifest from the figure that the upper bound of tan$\beta$ is given by
LSP condition. The figure vividly show the competition between
the lightest neutralino and light stau for the LSP position.
Another property worthy of mention is that most sparticles are insensitive
to tan$\beta$ when tan$\beta$ is large except $m_{\tilde{\tau}_1}$ $m_{A^0}$ and 
$m_{H^{\pm}}$.

\section{Constraints from $b \rightarrow s \gamma$}

It is well known that $b \rightarrow s \gamma$ put a very stringent constraint on
parameter space of various models. In this section we analyze the constraints from
$b \rightarrow s \gamma$ on the permitted parameter space discussed in the last
section. It is known long ago that supersymmetric contribution can interfere either
constructively or destructively\cite{LNWZ,BG,D,GO,GN},which is determined by the 
sign of $\mu$. 
For $\mu < 0$, the SUSY contribution interferes destructively with the Higgs's and
W's contributions. With the spectrum of sparticles low, both charged Higgs and
supersymmetric particles can largely contribute to the process. So even if the charged 
Higgs has large contributions, the supersymmetric contribution will cancel its effect
and, for large tan$\beta$, can even overwhelm its and W's contributions and force
the $C_7$ ($ C_7$ is the Wilson coefficient of the operator $O_7$ in the effective
Hamiltonian, eq.(36), and the branching ratio of $b\rightarrow s \gamma$ is determined
by $|C_7|^2$.) to change sign from positive to negative while keep
the branch ratio still safely stayed in the bounds of experiments.

As we pointed out in the last section, there exists a region where the mass spectrum
is low while tan$\beta$ is large. It is known that supersymmetric contribution
is proportional to tan$\beta$ in the region. So it is expected that in this region supersymmetric
contribution will be very large.

Figure 3 is devoted to show $b \rightarrow s \gamma$ constraint. The curves in figure 3a) which have a dip correspond to the upper limit of tan$\beta$, while the other
correspond to lower limit of tan$\beta$. The curves in figure 3b) which have a
convex correspond to the upper limit of tan$\beta$, while the other correspond the 
lower limit of tan$\beta$. The experimental 
bounds of $ b\rightarrow s \gamma$ are translated into the bounds of $C_7$. But it 
should be reminded that $C_7$ can be either negative or positive. So we map the allowed parameter space
into the plane of $M_{1/2}$ and $C_7$. Figure 3a) is for $\mu < 0$. It is apparent
that for $\Lambda_{SUSY}$= $2\times 10^{16}$ GeV  most of the allowed region of the 
quasi-massless scenario can safely pass the
experimental constraint due to the cancellation of the supersymmetric contribution to 
that of charged Higgs bosons when tan$\beta$ increases as shown by the line corresponding
to the upper limit of tan$\beta$, except for some part close to the small
tan$\beta$ boundary, where charged Higgs contributes much, while
supersymmetric particles contribute less since tan$\beta$ is small. For the dilaton 
dominant scenario and $\Lambda_{SUSY}$ equals $2\times 10^{16}$ GeV, we can see from 
figure 3a) that a quite large region is outside the experimental bound. This is 
because in the region SUSY contribution is not large enough to make $C_7$ still in
experimental bound after cancelling out contributions of
charged Higgs and W bosons. 
But for the case $\Lambda_{SUSY}$=$10^{14}$GeV, one can see from figure 3a) that there
is a region where supersymmetric contribution
indeed overwhelms charged Higgs's and W's contributions and makes
$C_7$ change its sign. In this region tan$\beta$ is large and the mass spectrum is low.
This region has interesting phenomenology which have been analyzed in Ref.~\cite{TT,HY,HLY,TT2,LW}
and we shall discuss in the next section. Recently, in order to make more precise 
theoretical prediction, many literatures are devoted to NLO corrections of this
process~\cite{N,C,KN}. If we include the NLO corrections the region will decrease because,
 as pointed out in Ref. \cite{C},  the NLO correction of supersymmetric contribution
decreases $30\%$, charged Higgs part decreases $20\%$, and SM part increases $10\%$.

Figure 3b) is devoted to $\mu > 0$. It is known that in such
case the supersymmetric contribution interferes constructively. So the low mass spectrum
region is not allowed whether tan$\beta$ is large or small for both scenarios.
One can see from the figure that for the quasi-massless
scenario, all parameter space allowed in section II can not give the
right prediction of $b \rightarrow s \gamma$ and is excluded by this strict
constraint. For the dilaton dominant scenario, as $M_{1/2}$ increases to 330 GeV, the
mass spectrum increases to such an extent that the region of low tan$\beta$ enters into
the experimental bound and is allowed . The typical
mass of sparticle is about 500 GeV in this region. 

For the dilaton dominant scenario and $\mu < 0$, it is possible to distinguish the 
interesting region where the mass spectrum is low and tan$\beta$ is large from the region
where the mass spectrum and tan$\beta$ both are large. We shall discuss the possibility
 in an analysis of the rare decay $b \rightarrow s \tau^+ \tau^-$.

\section{Some phenomenological predictions}

We now proceed to the analysis of low energy phenomenology. We shall discuss the rare
decay $b\rightarrow s \tau^+\tau^-$ and Higgs boson productions $e^+ e^-\rightarrow
b \bar{b} H$. In order to search significant SUSY effects we shall concentrate on the
case of the dilaton dominant scenario and $\mu < 0 $.

\subsection{Decay $b \rightarrow s \tau^+ \tau^-$}

The effective Hamiltonian relevant to 
the $b\rightarrow s l^{+}l^{-}$ process is 
\begin{eqnarray}
H_{eff} & = & \frac{4G_{F}}{\sqrt{2}} V_{tb} V_{ts}^{*} (\sum_{i=1}^{10}
    C_{i}(\mu) O_{i}(\mu) + \sum_{i=1}^{10} C_{Q_i}(\mu) Q_i(\mu))
\end{eqnarray}
where $O_i(i=1, 2, ..., 10)$ are given in Ref.~\cite{GSW}, and $Q_i$'s come from exchanging 
neutral Higgs bosons and have been given in Ref.~\cite{DHH}. The coefficients $C_i(m_w)$ 
and $C_{Q_i}(m_w)$ in SUSYMs have been calculated~\cite{BBMR,CMW,GOST,HY,HLY}. The branching ratio and
backward-forward (B-F) asymmetry for $b \rightarrow s \tau^+ \tau^-$ depend on the 
coefficients $C_7, C_8, C_9, C_{Q_1}$ and $C_{Q_2}$.

As pointed out in Ref.\cite{HY,HLY}, once $C_{Q_1}$ and $C_{Q_2}$ can complete with 
$C_8$ and $C_9$, both invariant mass
distribution and backward-forward asymmetry will be greatly modified.
The values of $C_{Q_1}$ and $C_{Q_2}$ depend on the mass splitting and the mixing 
angle of stops, the masses of charginos and diagonizing matrices $U$ and $V$, the 
masses of neutral Higgs bosons, and 
tan$^3\beta$ when tan$\beta$ is large. For small
masses of light chargino and neutral Higgs boson, large mass splitting of stops and 
large tan$\beta$,  $C_{Q_1}$ and $C_{Q_2}$ can be very large.

It is noted in the last section that, in the case of dilaton dominant scenario, 
$\mu < 0$ and $\Lambda_{SUSY}$=$10^{14}$ GeV,
after taking into account the constraint of $b \rightarrow s \gamma$,
there does exist a region (we shall call it the region A) of the parameter space where masses of sparticles are lower 
and tan$\beta$ can up to 25. In figure 4, we map the allowed parameter space into $C_{Q_1}$ and $M_{1/2}$
plane and $C_{Q_2}$ and $M_{1/2}$ plane respectively. The lower boundary of tan$\beta$ corresponds to the line near the
$M_{1/2}$-axis, while the upper boundary to the another line. It is obvious that the values of 
$C_{Q_1}$ and $C_{Q_2}$ indeed are very large in this region. We choose $M_{1/2}$=110 GeV and tan$\beta$
=23 as a representative point in the region and the values of $C_{Q_i}$ (i=1,2) as well 
as $C_i$ (i=7,8,9) at the point are tabulated in Table 1. It is also noticed in the
last section that there is another region (we shall call it the region B) in the allowed parameter space where the 
mass spectrum and tan$\beta$ both are large. In the region B, because tan$\beta$ can 
be up to 33,$C_{Q_1}$ and $C_{Q_2}$ can also compete with $C_8$ and $C_9$. But the 
values of $C_{Q_1}$ and $C_{Q_2}$ in this region are smaller when compared with those 
in the region A. In order to distinguish this region from the region A we have chosen
$M_{1/2}$=400 GeV and tan$\beta$=31 as a representative in this region to do 
calculations. The values of $C_{Q_i}$ (i=1,2) and $C_i$ (i=7,8,9) at the point are also
tabulated in Table 1. One can see from the Table that  a typical $C_{Q_1}$ in the 
region A and is -16, while a typical $C_{Q_1}$ in the region B is -4.5. Some masses of
sparticles used in computations are listed in Table 2.

The numerical results of the invariant mass distribution and B-F asymmetry for
the two sets of values of coefficients $C_{Q_i}$ and $C_i$ given in table 1 are shown
in figure 5. It is
obvious that the deviation from SM is very large for both cases, but for the set A
the deviation is more drastic. The enhancement of the differential branching ratio 
$d{\Gamma}/{ds}$ in the case of set A can reach $300\%$ compared to SM. Meanwhile,the 
deference between the set A and set B is also very significant so that one can 
distinguish them from the measurements of $b\rightarrow s \tau^+\tau^-$. It should be
noted that if without including the contributions of neutral Higgs, the deviation from
SM is small.

\begin{table}[htbp]
\begin{center}
\begin{tabular}{|c|c|c|c|c|c|c|c|c|}
\hline
 & $M_{1/2}$ & tan$\beta$ & $C_7$ & $C_8$ & $C_9$ & $C_{Q_1}$ & $C_{Q_2}$ & BR($b\rightarrow s \gamma$)\\
 \hline
Set A & 110 & 23 & -0.25 & -3.08 & 4.12 & -16.64 & 16.36 & $2.14 \times 10^{-4}$\\
\hline
Set B & 400 & 31 & 0.24 & -3.06 & 4.50 & -4.35 & 4.30 & $2.0 \times 10^{-4}$\\
 \hline
  \end{tabular}
 \end{center}
\caption{The values of $C_{Q_i}$ (i=1,2) and $C_i$ (i=7,8,9) for the chosen 
representative points in the regions A and B. }
\vspace{1cm}
 \end{table}
\begin{table}[htbp]
\begin{center}
\begin{tabular}{|c|c|c|c|c|c|c|c|c|c|}
\hline
 & $m_{\tilde q}$ & $m_{\tilde {t}_1}$ & $m_{\tilde {t}_2}$ & $m_{\chi_1}$ & $m_{
\chi_2}$ & $m_{h_0}$ & $m_{H^{\pm}}$ & $m_{\tilde {\tau}_1}$ & $m_{\chi^0_1}$\\
 \hline
Set A & 246.30 & 162.95 & 336.86 & 73.58 & 192.56 & 103.13 & 153.31 & 50.18 & 50.16\\
\hline
Set B & 797.60 & 575.75 & 784.00 & 332.08 & 512.48 & 116.68 & 464.21 & 206.67 & 206.60\\
 \hline
  \end{tabular}
 \end{center}
\caption{The masses of sparticles used in the computations 
for the chosen representative points in the regions A and B. }
\label{Table 2}
\vspace{1cm}
 \end{table}

\subsection{$e^+e^-\rightarrow b\bar{b}H$}

The Higgs boson is the missing piece and also the least known one of   
the standard model and other supersymmetrical models. 
 The pursuit of the Higgs   
bosons predicted by these models is one of the primary goals of the present   
and next generation of colliders.   
The Next Linear Collider(NLC) operating at a center-of-mass energy of $   
500-2000 GeV$ with the luminosity of the order of $10^{33} cm^{-2} s^{-1}$   
can provide an ideal place to search for the Higgs boson, 
since the events would be much cleaner than in the LHC   
and the parameters of the Higgs boson would be easier to   
extracted.  

Based on the analysis in previous sections, we will present some data examples
of the cross sections for the process 
$e^+e^- \rightarrow b\bar b H$ in this subsection. 
In figure 6
we show the SM Higgs production cross section
as a function of the Higgs mass. In figure 7 and 8, 
we show the production cross sections as the
function of the $\tan\beta$, where $M_{1/2}=120$ GeV and $400$ GeV,
respectively, and other parameters are depicted in figures captions. As usual, $h^0$
and $H^0$ denote the CP-even neutral Higgs bosons with $m_{h^0}$ $<$ $m_{H^0}$, 
respectively. It should be noticed that, as $M_{1/2}=400$ GeV, 
the mass of $H^0$ is too heavy to be produced by the NLC when $\sqrt{s}=500$ GeV.
Comparing figures 7 and 8 with figure 6, it is evident that Higgs production cross
sections increase significantly when tan$\beta$ increases, as expected, except for
$b\bar b h^0$ in the case of $M_{1/2}$=400 GeV. For the $b\bar b h^0$ production,
the enhancement of large tan$\beta$ is offseted by the small sin$\alpha$ because
in the case of $M_{1/2}$ = 400 GeV, $m_{h^0}$ is much smaller than masses of the other
Higgs bosons.

\section{Conclusions}

We have revisited the the eleven dimension Planck scale, the
physical scale of the eleventh dimension, the physical scale of  Calabi-Yau manifold 
 and coupling in hidden sector in M-theory on $S^1/Z_2$ and discussed the reasonable
bounds on them under the ansatz that the scale of $M_{GUT}$, $M_H$ and 
$\left[\pi \rho_p\right]^{-1}$  should be lower than the eleven dimension Planck scale. 
It has been shown that $\Lambda_{SUSY}\geq 10^{14}$ GeV if one assumes $\Lambda_{SUSY}$
=$[\pi \rho_p]^{-1}$~\cite{LLN}. Choosing 2 representative points
which correspond to scalar quasi-massless scenario and
dilaton dominant SUSY breaking scenario
respectively, we have calculated 
sparticle spectra at 
different values of $\Lambda_{SUSY}$ and found that the spectra lower when 
$\Lambda_{SUSY}$ decrease. Therefore, compared
with the spectra in the weakly coupling string models and general SUSY GUT models, the
spectra in M theory phenomenology are lower, which is, of course, more easier to 
search at colliders. The LEP experiment and $b\rightarrow s\gamma$ constraints on the
parameter space in the dilaton dominant and scalar quasi-massless supersymmetry
breaking scenarios are analyzed. Finally, we give predictions for the rare decay $b
\rightarrow s\tau^+\tau^-$ and neutral Higgs boson productions. An interesting result is
that one could discover supersymmetry from $b\rightarrow s\tau^+\tau^-$ in B factories
if nature give us large tan$\beta$ and low mass spectra which come out as a 
consequence of M theory low energy phenomenology.
\section{Acknowledgements}
This research was supported in part by the National Science
Foundation of China. This research ( T. Li ) was supported in part by the 
U.S.~Department of Energy under
Grant No.~DE-FG02-95ER40896 and in part by the University 
of Wisconsin Research Committee with funds granted by the 
Wisconsin Alumni Research Foundation.
This work was also supported in part by the post doctoral foundation
of China and the author gratefully acknowledges the
support of  K.C. Wong Education Foundation, Hong Kong.

\vskip 1cm
\begin{center}
{\Large \bf Figure Captions}
\end{center}
\vskip 0.5cm
\noindent Fig.\ 1:

The upper and lower bounds of tan$\beta$ vary with $M_{1/2}$, for the dilaton 
dominant and scalar quasi-massless scenarios and for different values of $\Lambda_{SUSY}$.
The curve lebeled 1(2) represents the scalar quasi-massless (dilaton dominant) scenario. 
The dotted line is for $\Lambda_{SUSY}$ = $10^{14}$ GeV, the dashed line $10^{15}$ GeV, and the
solid line the GUT scale, $2 \times 10^{16}$ GeV. For the scalar quasi-massless
scenario, the permitted parameter spaces are with closed boundaries, while for the 
dilaton dominant scenario, the permitted parameter spaces are not closed in right parts. 
Fig.\ 1a) is for $\mu <0$, fig.\ 1b) is for $\mu>0$.
The lower bound of tan$\beta$ is about 1.6 for the dilaton dominant scenario.
\vskip 0.5cm
\noindent Fig.\ 2:

Computed values of super-particle masses versus tan$\beta$ for
$M_{1/2}$=120 GeV in dilaton scenario.
Fig.\ 2a) is for $\Lambda_{SUSY}= 10^{14}$ GeV and Fig.\ 2b) $1\times 10^{16}$ GeV.
The solid (dashed) lines represent $\mu > (<) 0 $.

\vskip 0.5cm
\noindent Fig.\ 3:

The variation of $C_7$ with $M_{1/2}$ and tan$\beta$, scenarios and $\Lambda_{SUSY}$. 
The solid lines represent $\Lambda_{SUSY}$ = $2 \times 10^{16}$ GeV, the dotted 
lines represent $\Lambda_{SUSY}$ = $1\times 10^{14}$ GeV. The curve lebeled 1
(2) represents the scalar quasi-massless (dilaton dominant) scenario. 
Fig.\ 3a) is devoted for the case $\mu <0$, fig.\ 3b) for the case $\mu>0$.
The experimental constraints of $b \rightarrow s \gamma$ have been translated
to the constraints on $C_7$, which are represented by two sets of parallel horizontal
lines. The curves in figure 3a) which have a dip correspond to the upper limit of tan$\beta$, while the other
correspond to lower limit of tan$\beta$. The curves in figure 3b) which have a
convex correspond to the upper limit of tan$\beta$, while the other correspond the 
lower limit of tan$\beta$.
\vskip 0.5cm
\noindent Fig.\ 4:

For the case of the dilaton dominant scenario, $\mu < 0$ and $\Lambda_{SUSY}$=$10^{14}$ GeV,
the variation of $C_{Q_1}$ and $C_{Q_2}$ with $M_{1/2}$ and tan$\beta$.
We map the permitted parameter space plane into $C_{Q_i}$ and $M_{1/2}$
planes. The lower boundary of tan$\beta$ corresponds to the line near the
$M_{1/2}$-axis, while the upper boundary to the other line.
\vskip 0.5cm
\noindent Fig.\ 5:

The invariant mass distribution of dilepton and B-F asymmetry for the process $b\rightarrow
s \tau^+ \tau^-$. The related coefficients and masses are listed in table 1 and table 2 
respectively. The solid line is for SM prediction, the dashed line corresponding to
the prediction of group A, the dotted line the prediction of group B. 
We find if without including the contributions of neutral Higgs, the deviation from
SM is small.
\vskip 0.5cm
\noindent Fig.\ 6:

The cross sections of the process $e^+e^-\rightarrow b \bar b H$ 
as a function of the mass of Standard Model Higgs $m_H$.
\vskip 0.5cm
\noindent Fig.\ 7:

The cross sections of the process $e^+e^-\rightarrow b \bar b H$($H=h^0, H^0$) 
as a function of $\tan\beta$, where $M_{1/2}=120$ GeV,
$\Lambda_{SUSY}=10^{14}$ GeV and $Sign(\mu)=-1$ in the dilaton dominant scenario.

\vskip 0.5cm
\noindent Fig.\ 8:

Same with figure 7 but $M_{1/2}=400$ GeV.

\newpage
\begin{figure}[t]
\vspace{0cm}
\centerline{
\epsfig{figure=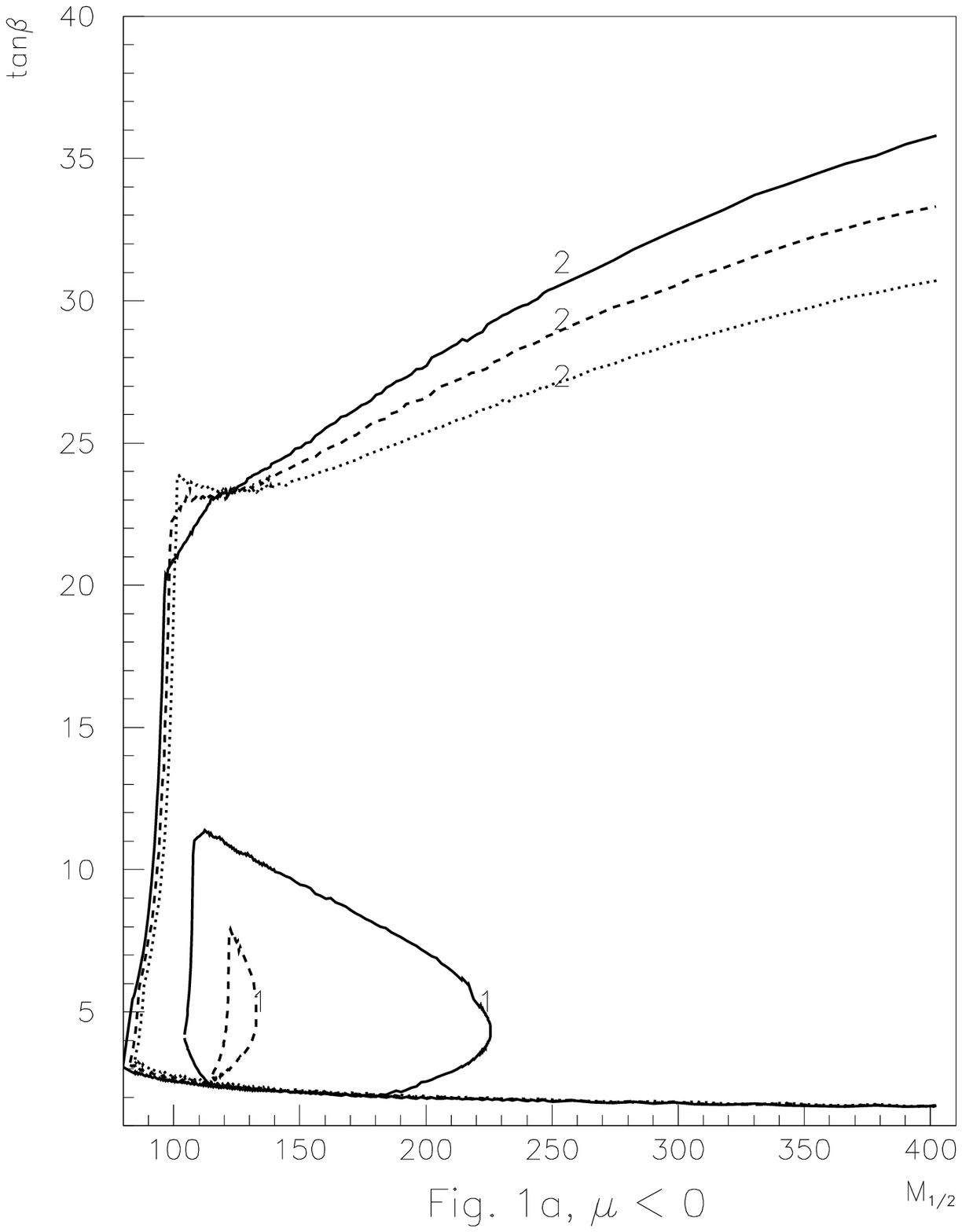,width=6 in}
}
\label{fig1a}
\end{figure}

\newpage
\begin{figure}[t]
\vspace{0cm}
\centerline{
\epsfig{figure=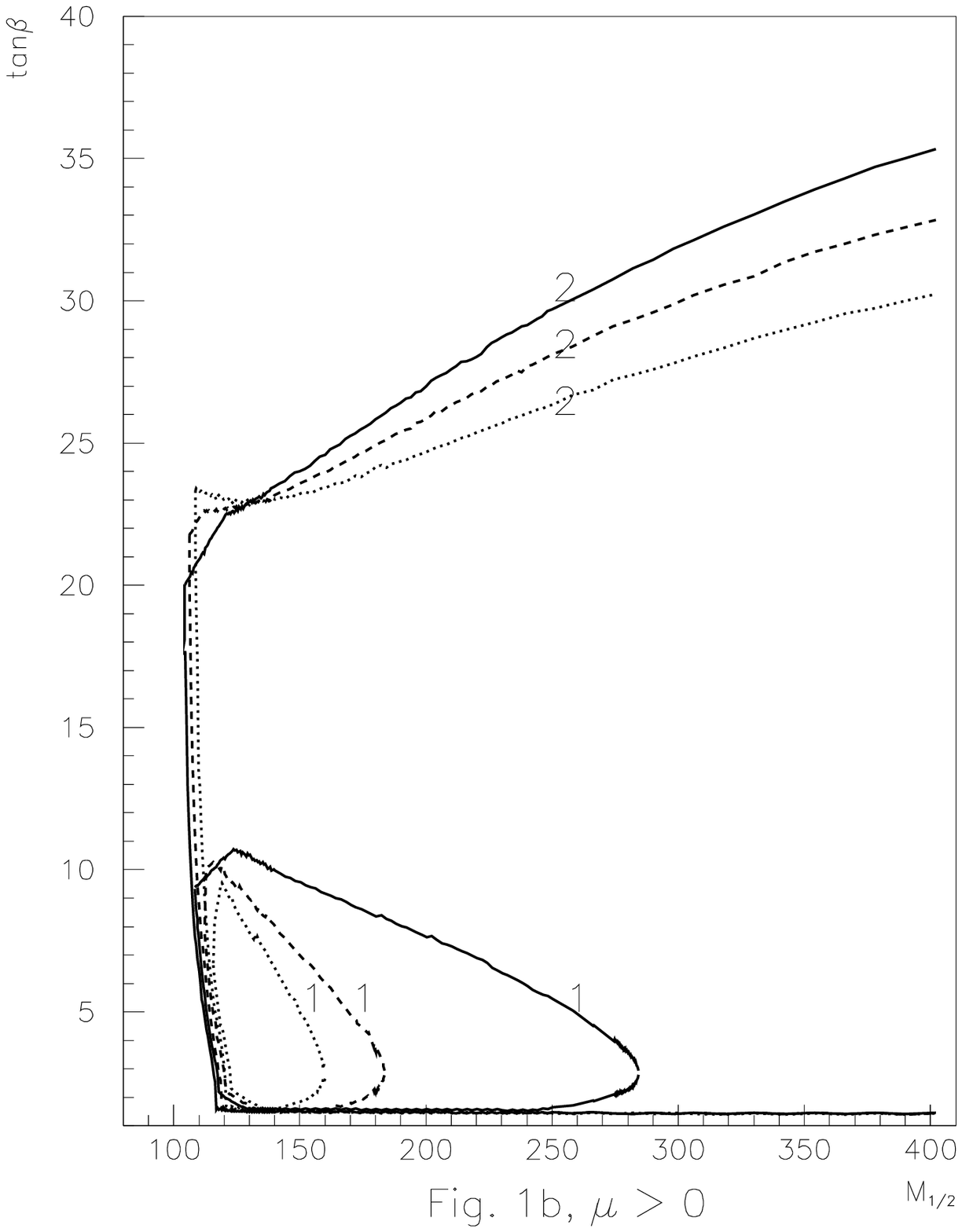,width=6 in}
}
\label{fig1b}
\end{figure}

\newpage
\begin{figure}[t]
\vspace{0cm}
\centerline{
\epsfig{figure=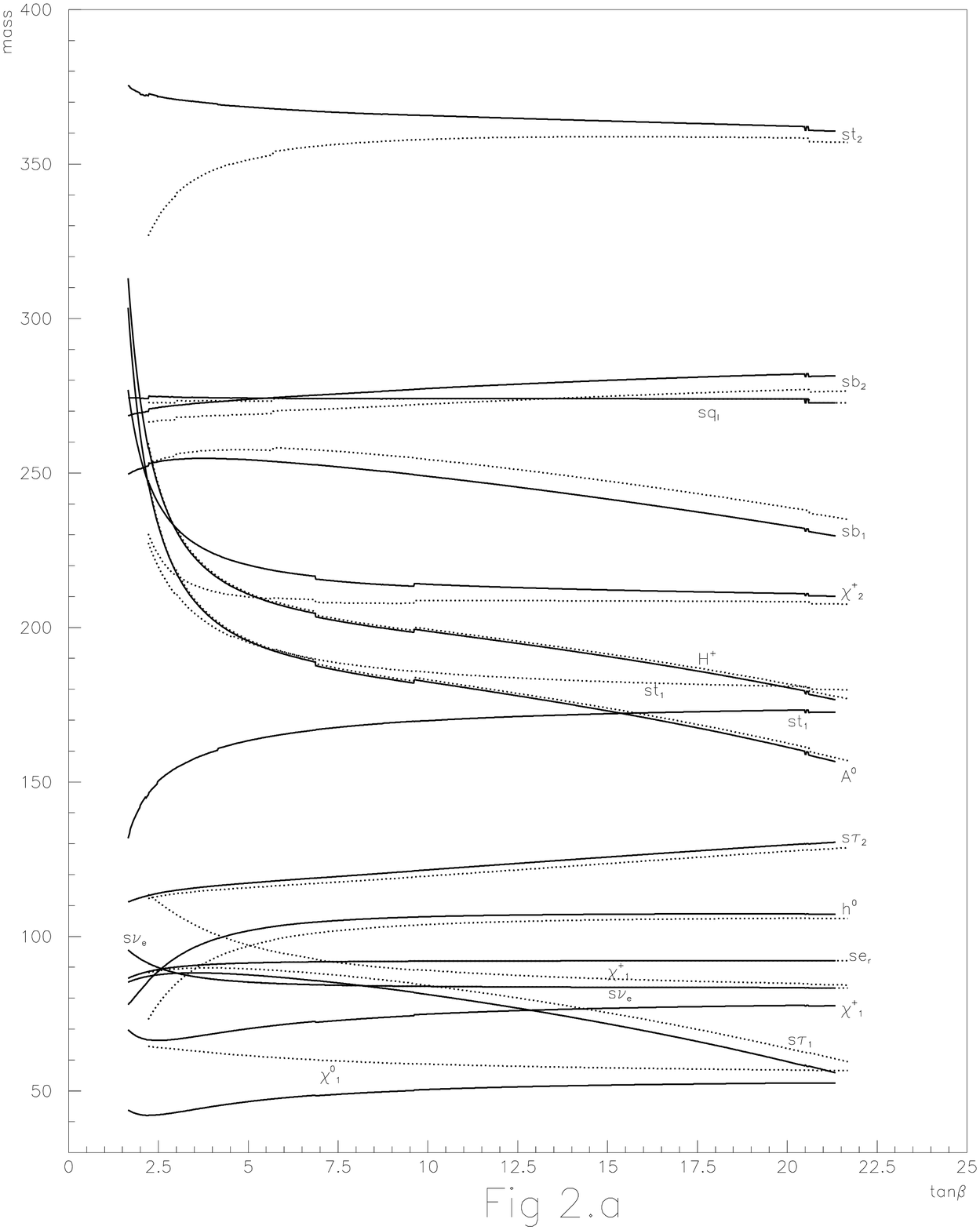,width=6 in}
}
\label{fig2a}
\end{figure}

\newpage
\begin{figure}[t]
\vspace{0cm}
\centerline{
\epsfig{figure=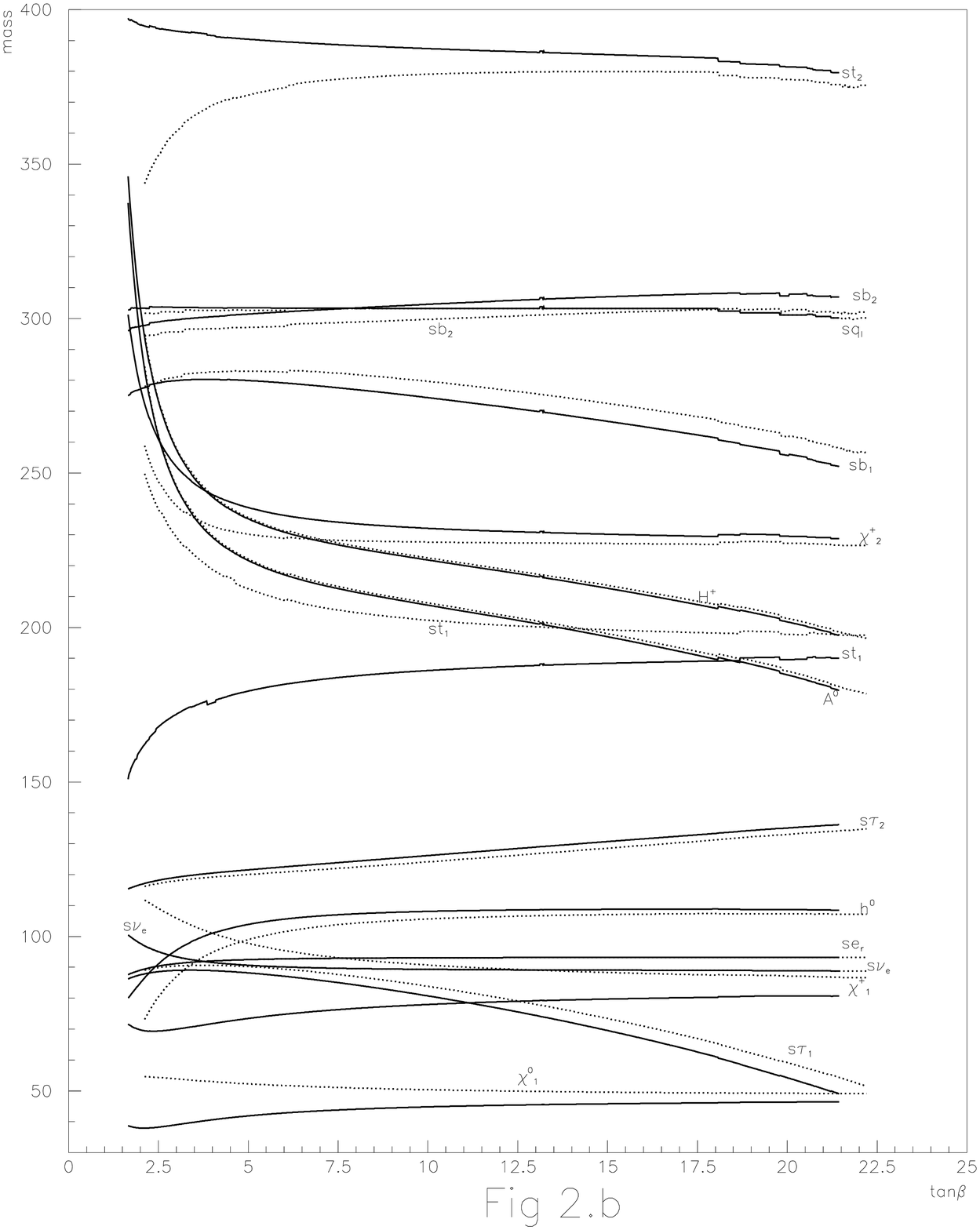,width=6 in}
}
\label{fig2b}
\end{figure}

\newpage
\begin{figure}[t]
\vspace{0cm}
\centerline{
\epsfig{figure=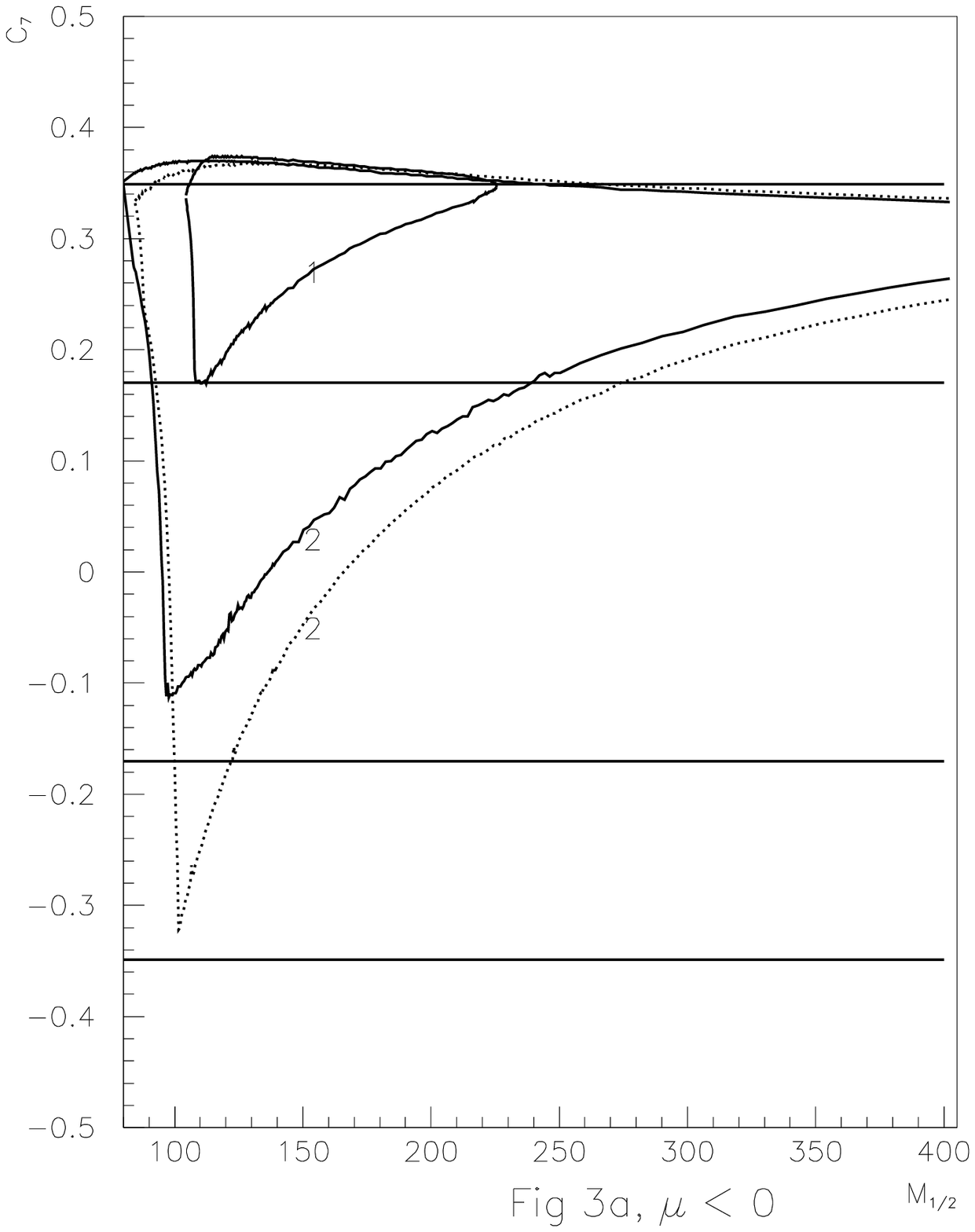,width=6 in}
}
\label{fig3a}
\end{figure}

\newpage
\begin{figure}[t]
\vspace{0cm}
\centerline{
\epsfig{figure=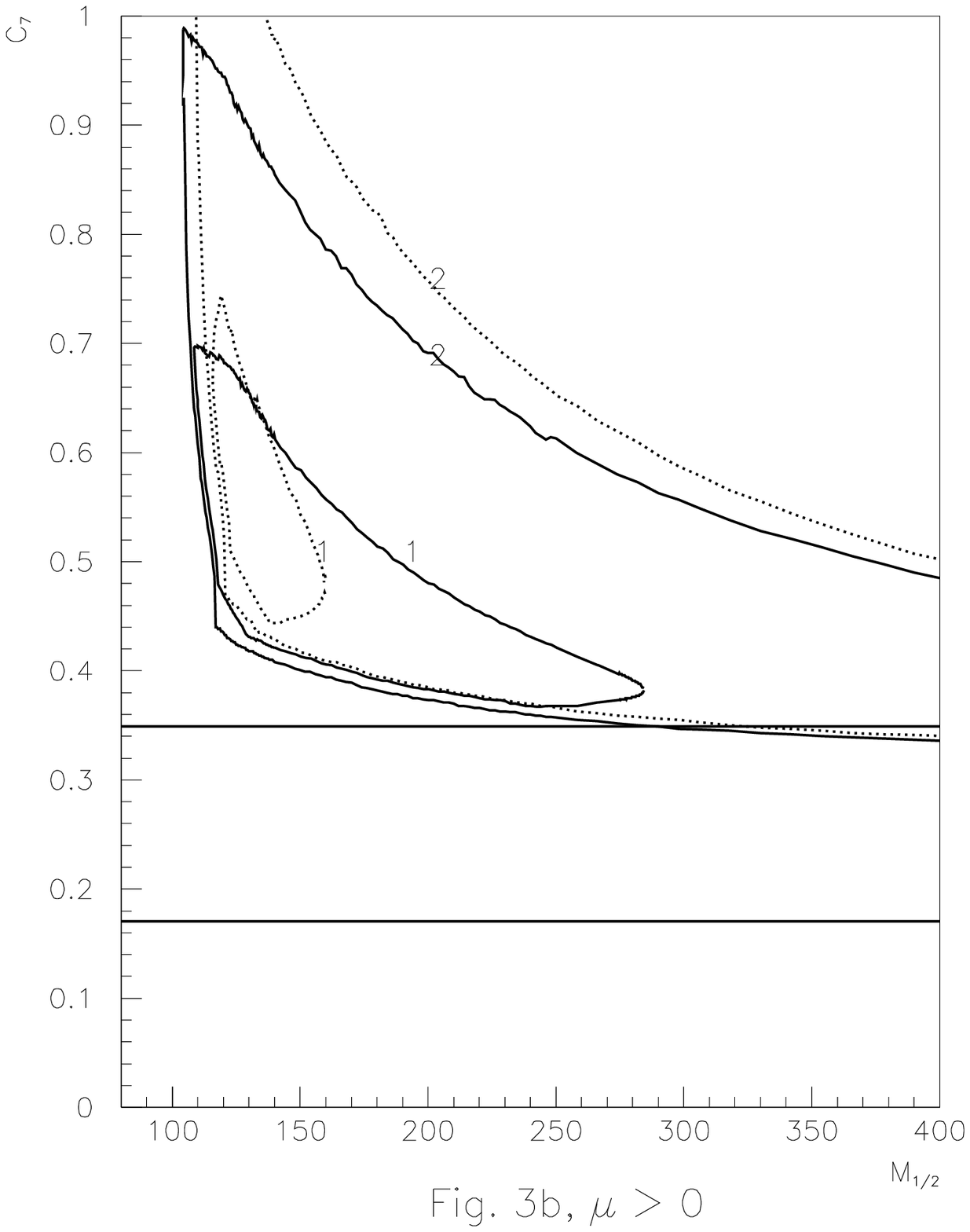,width=6 in}
}
\label{fig3b}
\end{figure}

\newpage
\begin{figure}[t]
\vspace{0cm}
\centerline{
\epsfig{figure=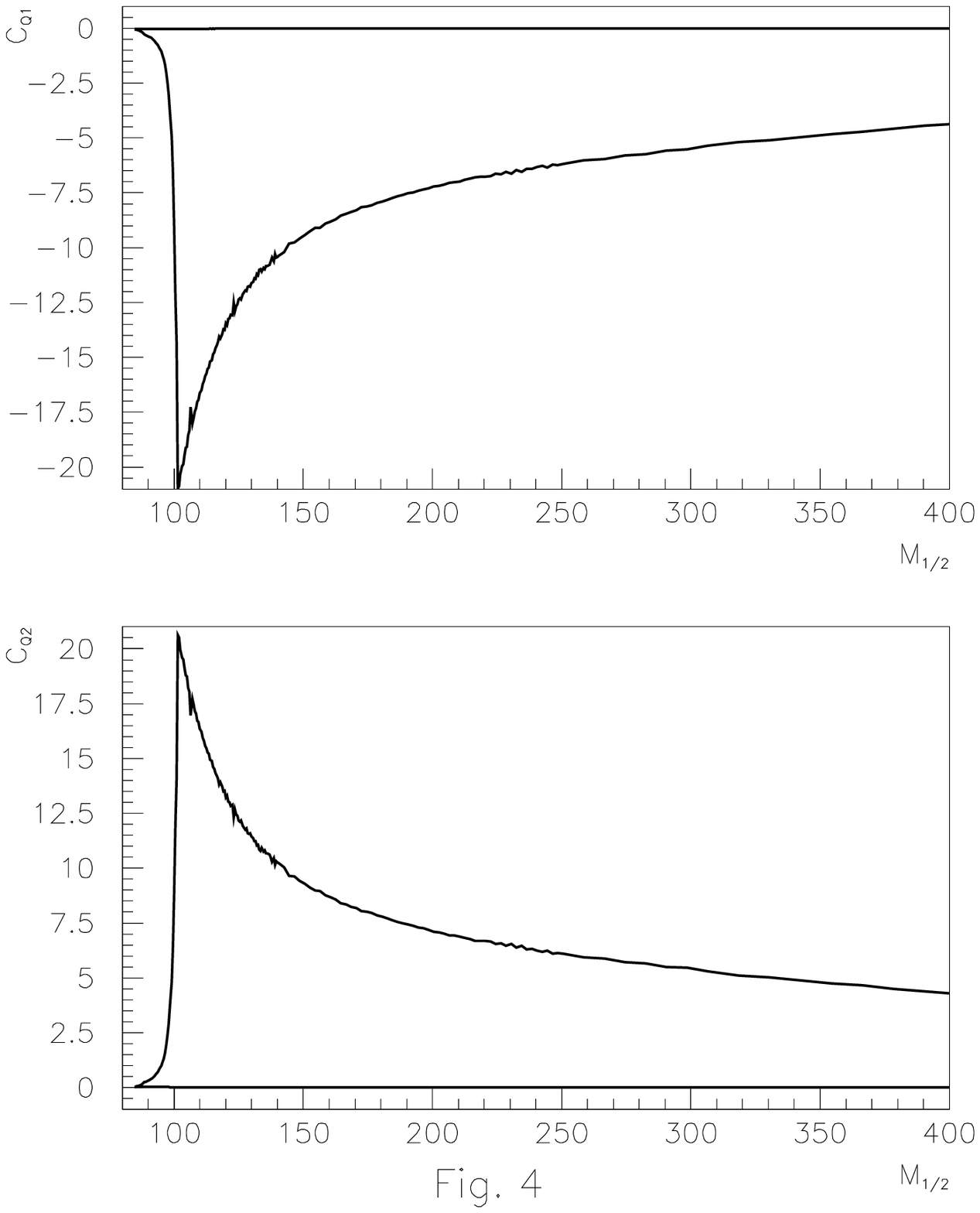,width=6 in}
}
\label{fig4}
\end{figure}

\newpage
\begin{figure}[t]
\vspace{0cm}
\centerline{
\epsfig{figure=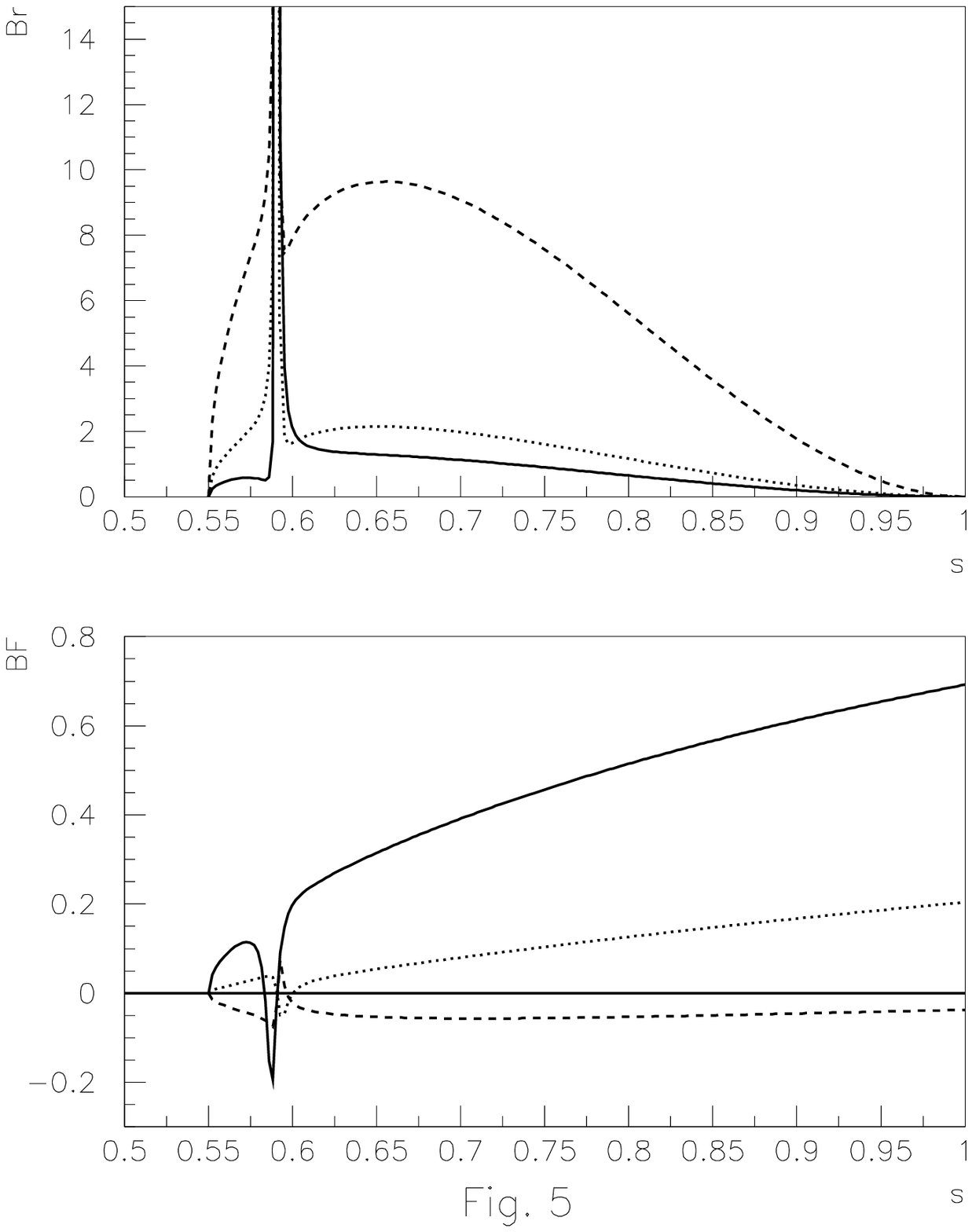,width=6 in}
}
\label{fig5}
\end{figure}

\newpage
\begin{figure}[t]
\vspace{0cm}
\centerline{
\epsfig{figure=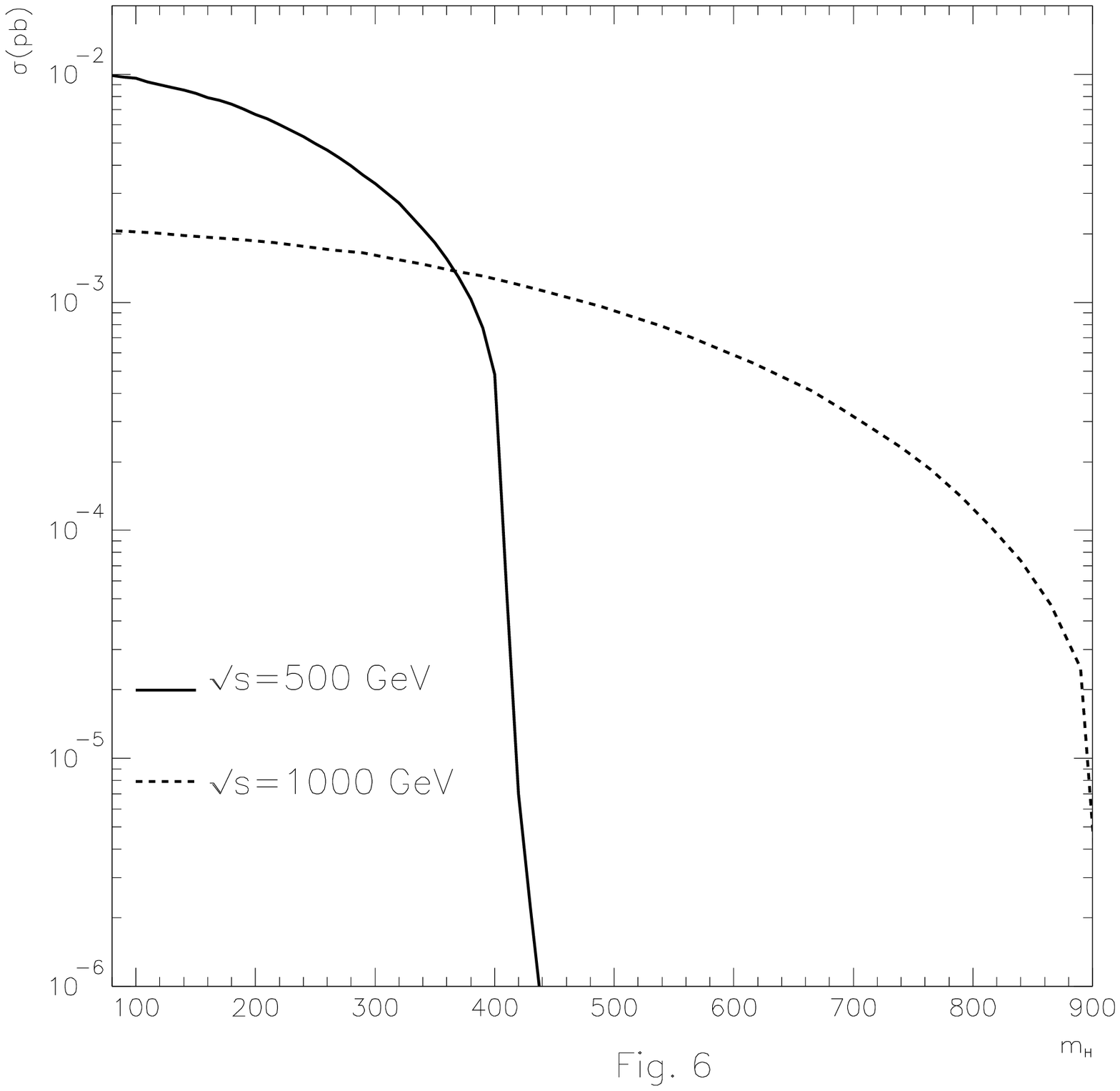,width=6 in}
}
\label{fig6}
\end{figure}

\newpage
\begin{figure}[t]
\vspace{0cm}
\centerline{
\epsfig{figure=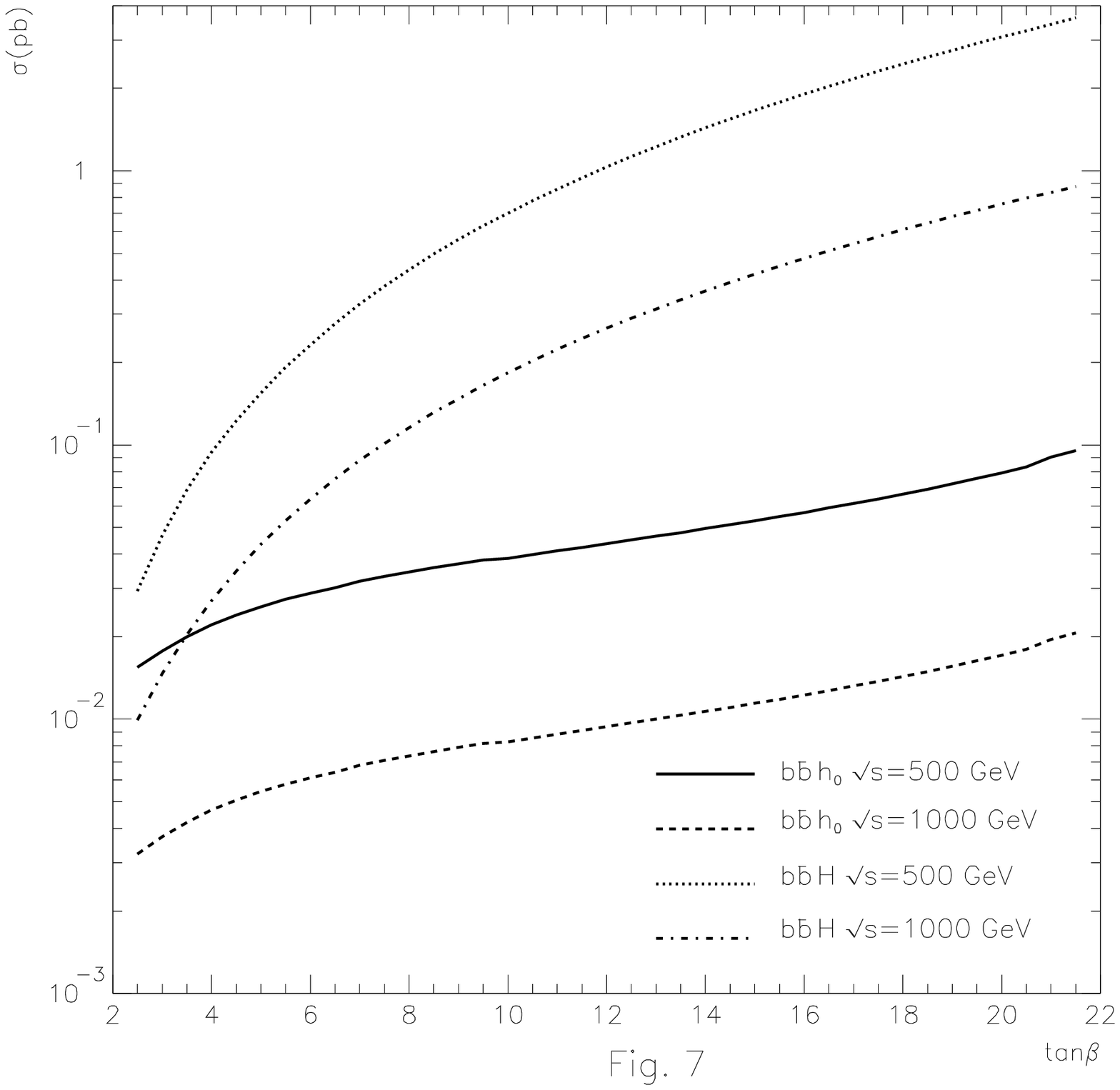,width=6 in}
}
\label{fig7}
\end{figure}

\newpage
\begin{figure}[t]
\vspace{0cm}
\centerline{
\epsfig{figure=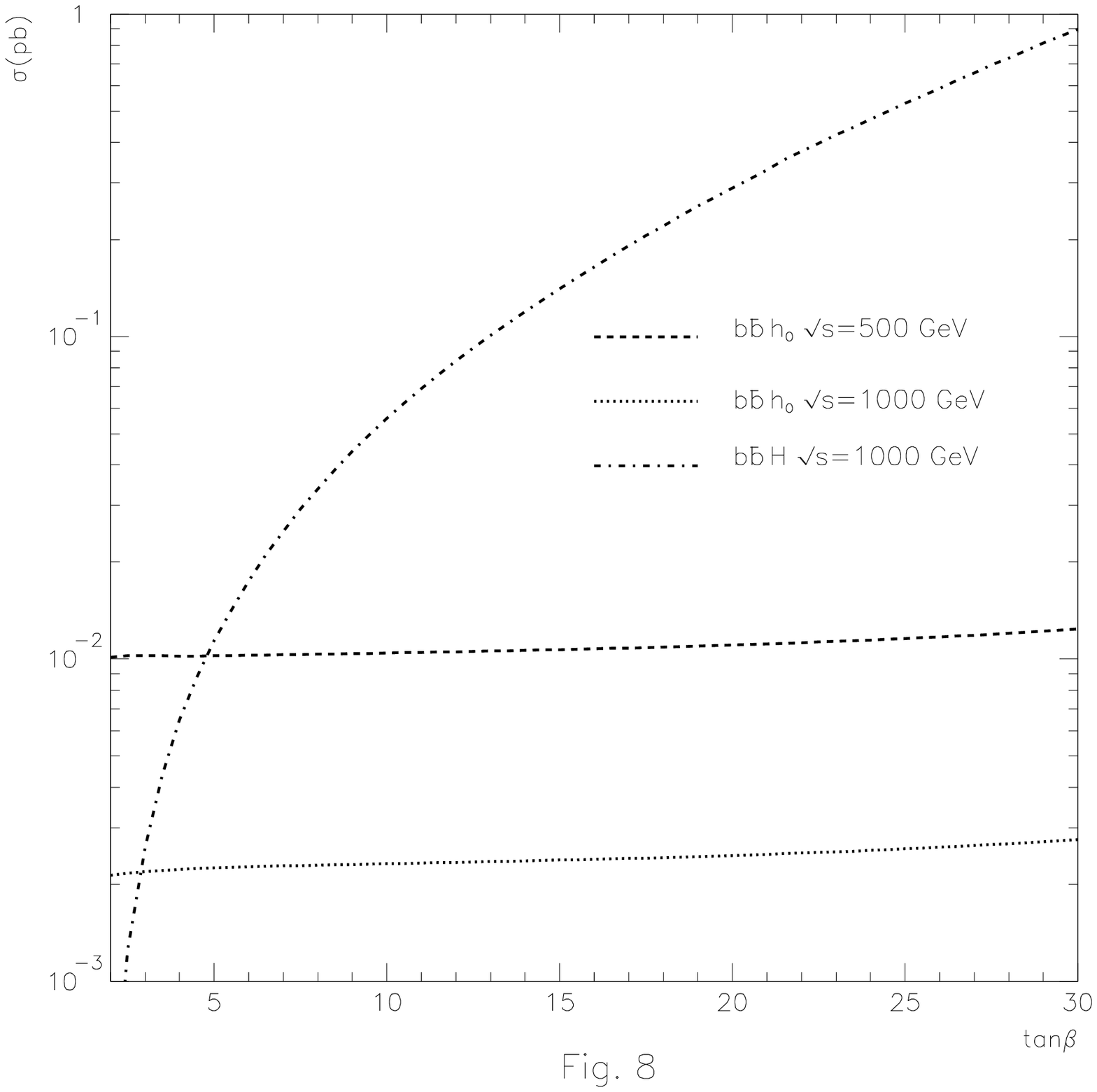,width=6 in}
}
\label{fig8}
\end{figure}
\end{document}